\documentclass[fleqn,usenatbib]{mnras}
\usepackage{newtxtext,newtxmath}
\usepackage[T1]{fontenc}
\usepackage{ulem} 

\DeclareRobustCommand{\VAN}[3]{#2}
\let\VANthebibliography\thebibliography
\def\thebibliography{\DeclareRobustCommand{\VAN}[3]{##3}\VANthebibliography}

\usepackage{graphicx}	
\usepackage{amsmath}	

\newcommand{\highlight}[1]{\textcolor{black}{\textnormal{{#1}}}}

\newcommand{\teff}[0]{$T_{\text{eff}}$}
\newcommand{\msun}[0]{M$_{\odot}$}
\newcommand{\logg}{\mbox{$\log g$}}
\newcommand{\vsini}{\mbox{$v \text{sin} i$}}
\newcommand{\logrhk}{\mbox{$\log R^{\prime}_{\text{HK}}$}}
\newcommand{\alpham}{\mbox{[$\rm \alpha$/M]}}
\newcommand{\yar}{\mbox{young $\alpha$--rich}}

\newcommand{\halpha}{\mbox{H$_{\alpha}$}}
\newcommand{\primerhk}{\mbox{$R^{\prime}_{\textrm{HK}}$}}
\newcommand{\rhk}{\mbox{$R_{\textrm{HK}}$}}
\newcommand{\rhkphot}{\mbox{$R_{\textrm{HK,phot}}$}}

\title[Fast rotation and strong activity in YAR stars]{\centering{New Evidence of Binarity in Young $\alpha$--Rich Turn-off and Subgiant Stars: Fast Rotation and Strong Magnetic Activity}}

\author[Yu et al.]{
Jie Yu,$^{1,2,3}$\thanks{E-mail: jie.yu@anu.edu.au (JY)}
Luca Casagrande,$^{2,3}$
Ioana Ciuc\u{a},$^{1,2,3}$
Yuan-Sen Ting,$^{1,2,4,5}$
Simon J. Murphy,$^{6}$
Boquan Chen$^{2,3}$
\\
$^{1}$School of Computing, Australian National University, Acton, ACT 2601, Australia\\
$^{2}$Research School of Astronomy \& Astrophysics, Australian National University, Cotter Rd., Weston, ACT 2611, Australia\\
$^{3}$ARC Centre of Excellence for All Sky Astrophysics in 3 Dimensions (ASTRO 3D), Stromlo, Australia\\
$^{4}$Department of Astronomy, The Ohio State University, Columbus, OH 43210, USA\\
$^{5}$Center for Cosmology and AstroParticle Physics (CCAPP),
The Ohio State University, Columbus, OH 43210, USA\\
$^{6}$Centre for Astrophysics, University of Southern Queensland, Toowoomba, QLD 4350, Australia}

\date{Accepted XXX. Received YYY; in original form ZZZ}

\pubyear{2015}

\begin{document}
\label{firstpage}
\pagerange{\pageref{firstpage}--\pageref{lastpage}}
\maketitle

\begin{abstract}
Young $\alpha$--rich (YAR) stars within the old Galactic thick disk exhibit a dual characteristic of relative youth determined with asteroseismology and abundance enhancement in $\alpha$ elements measured from high--resolution spectroscopy. The youth origin of YAR stars has been proposed to be binary evolution via mass transfer or stellar mergers. If that is the case, YAR stars should spin rapidly and thus be magnetically active, because they are mass and angular momentum gainers. In this study, to seek this binary footprint we select YAR stars on the main--sequence turn--off or the subgiant branch (MSTO--SGB) from APOGEE DR17, whose ages and projected rotation velocities (\vsini) can be precisely measured. With APOGEE \vsini\ and LAMOST spectra, we find that YAR stars are indeed fast rotators and magnetically active. In addition, we observe low [C/N] ratios and high Gaia RUWE in some YAR stars, suggesting that these MSTO--SGB stars probably have experienced mass transfer from red--giant companions. \highlight{Our findings underscore that magnetic activity can serve as a valuable tool for probing the binary evolution} for other chemically peculiar stars, such as red giants with lithium anomalies and carbon--enhanced metal--poor stars.
\end{abstract}




\section{Introduction}
Approximately a decade ago, a small sample of red giants exhibiting enhanced $\alpha$ abundances in the old, thick disk of the Galaxy were found to be young with the aid of asteroseismology, dubbed \yar\ (YAR) stars \citep{chiappini2015, martig2015}. \highlight{This led to a revised interest to earlier discoveries of stars with unusual ages for their chemistry \citep[e.g.,][]{fuhrmann1999, fuhrmann2011}.} Since YAR stars are outliers in the age-\alpham{} relation of our Galaxy \highlight{\citep{haywood2013, chiappini2015}}, binary evolution in the form of mass transfer or stellar mergers has been then \highlight{discussed as the youth origin \citep{chiappini2015, martig2015}}. In this scenario, YAR stars are mass gainers, and thus merely appear to be young with respect to isochrones due to their current high--mass nature, while kinematic data suggest that they are potentially old stars, consistent with their neighbours in the thick disk \highlight{\citep[e.g.,][]{aguirre2018, miglio2021, sun2020, ciuca2021, zhang2021a, cerqui2023}}. As such, radial velocity (RV) surveys have been conducted to search for the large RV variations expected from binary orbits of those stars \citep{yong2016, jofre2016, matsuno2018, jofre2023}. Moreover, Gaia RUWE data \citep{lindegren2018}, which serve as a robust metric for testing for unresolved multiplicity of Gaia sources \citep{belokurov2020}, have been used to trace binarity in YAR stars \citep{jofre2023}. In addition, simulations and observations have showed that [C/N] ratios may be altered by binary evolution \citep[e.g.,][]{izzard2018, hekker2019, zhang2021a, jofre2023, grisoni2023}. Differing across these three metrics, our objective in this study is to use a novel approach to seek robust evidence in support of binary evolution in YAR stars.

\begin{figure*}
\includegraphics[width=1.0\textwidth]{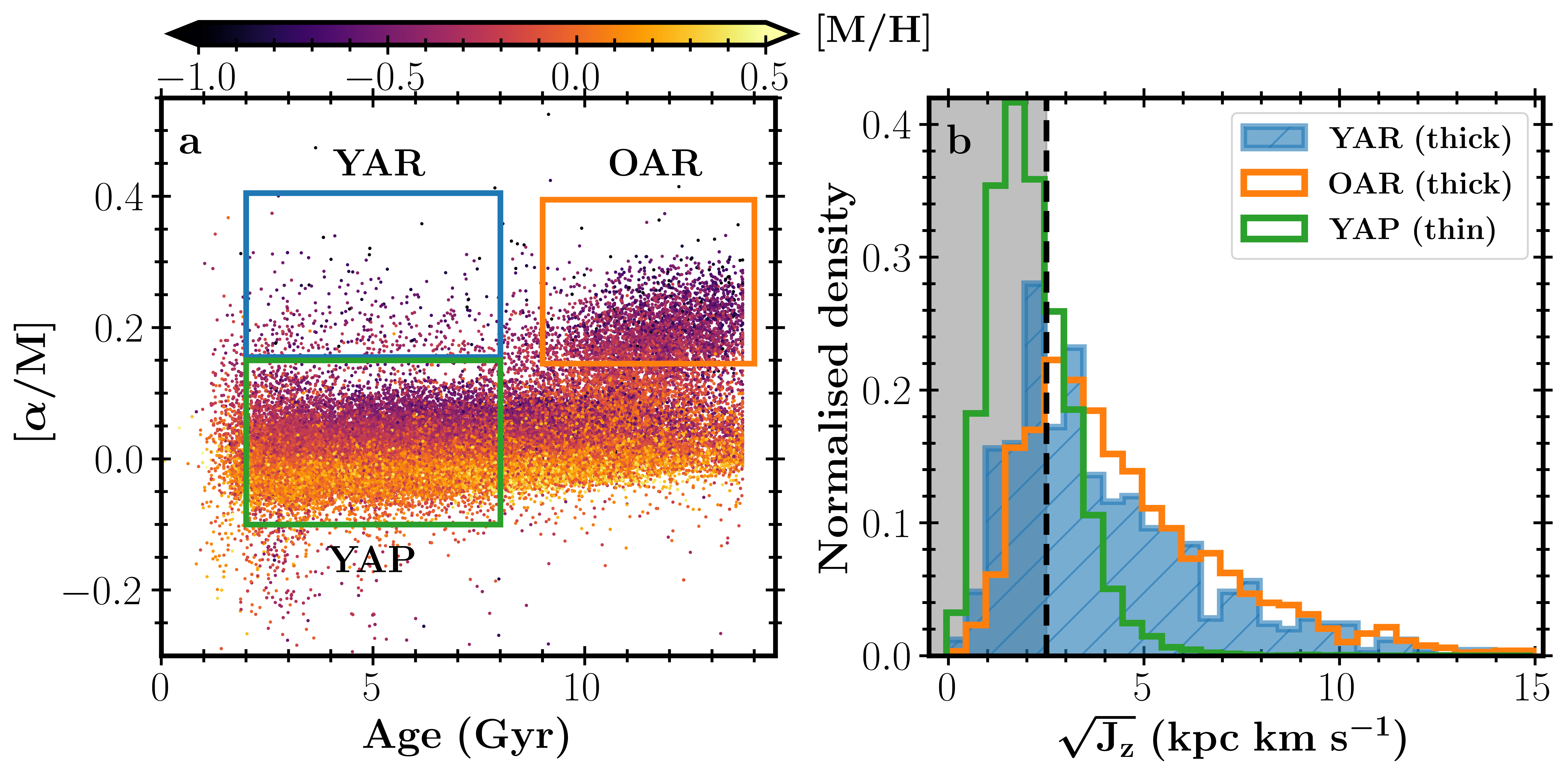}
\caption{Target selection of young $\alpha$--rich (YAR) stars as well as two reference samples of old $\alpha$--rich (OAR) stars and young $\alpha$--poor (YAP) stars. \textbf{Panel a}: Ages and [$\alpha$/M] cuts, as indicated in the three boxes (also see the text). Chemical abundances of $\alpha$ elements and overall metallicities are adopted from APOGEE DR17 and stellar ages from the \texttt{StarHorse} catalog based on isochrone fitting. All the stars shown here are in the evolutionary phase of the MSTO--SGB. \textbf{Panel b}: Kinematics cuts. The YAR (blue) and OAR (orange) samples are further refined by restricting the square root of their vertical actions to $>$~2.5~kpc~km~s$^{-1}$ (the region without grey shading), as indicated by the vertical dashed line.
}\label{fig:sampleselection}
\end{figure*}

The binary evolution hypothesis motivated us to investigate the rotation and magnetic activity of YAR stars. Suppose YAR stars are mass gainers and as such angular momentum gainers, one would expect YAR stars to be fast rotators, and hence magnetically active due to the correlation between stellar rotation and activity \citep[see][for a recent review]{isik2023}, compared with other stars in the old, thick disk. Notably, we emphasize that in certain cases, stellar rotation and magnetic activity may serve as superior metrics compared with RV variability and Gaia RUWE, particularly in scenarios involving mergers. This is because the former two parameters are detectable over stellar evolutionary timescales \citep{santos2019, santos2021, reinhold2020, zhang2020}. In contrast, the latter two pertain to binary orbits and become challenging to detect because mergers occur on short, dynamical timescales \citep[see][and references therein]{ivanova2020, henneco2023}.

In this work, we focused on the YAR stars on the main--sequence turn--off or the subgiant branch (MSTO--SGB) rather than red giants for three reasons. First, their ages can be measured with precision through isochrone--fitting methods, due to the correlation between their ages and luminosities \citep{queiroz2023}. We note that while asteroseismology is a benchmark for stellar ages, known YAR stars exhibiting solar--like oscillations remain limited in number \citep{hekker2019, jofre2023, grisoni2023}. Second, projected surface rotation rates, \vsini, are more readily measurable with spectral lines for MSTO--SGB stars, which is usually not the case for red giants that have been widely used to study YAR phenomena. 
Third, the [C/N] ratios of MSTO--SGB stars remain uncontaminated by the first dredge--up along the red giant branch, if binary interaction is absent. Therefore, observations of an excess of YAR stars with exceptionally low [C/N] ratios can be linked to the influence of the binary interactions associated with red giants.

\section{Sample Selection} \label{sec:sample}
To compile a population of young $\alpha$--rich stars in the evolutionary phase of the MSTO--SGB, we started with the APOGEE DR17 sample contained in the \texttt{StarHorse} catalog \citep{queiroz2023}, which provides stellar ages based on isochrone fitting. There, \mbox{MSTO--SGB} stars were selected with \teff\ and \logg\ cuts (see their equations 1 and 2 and figure 5). Among these stars, we selected a YAR star sample having stellar ages ranging from 2 to 8 Gyr and [$\rm \alpha$/M] ratios between 0.15 and 0.40 (see Fig.~\ref{fig:sampleselection}a), where the [$\rm \alpha$/M] ratios were retrieved from APOGEE DR17 \citep{abdurrouf2022}. For comparison, we defined two reference samples: young $\alpha$--poor (YAP) stars and old $\alpha$--rich (OAR) stars. We selected YAP stars by requiring $2<\text{age/Gyr}<8$ and $-0.10<[\rm \alpha/M]<0.15$, and OAR stars by demanding $9<\text{age/Gyr}<14$ and $0.15<[\rm \alpha/M]<0.40$ (see Fig.~\ref{fig:sampleselection}a). The lower age limit at 2 Gyr was set to ensure high age accuracy, as suggested by \citet{queiroz2023}. We retained stars with unset \texttt{STAR\_BAD} flags in APOGEE DR17 to ensure high accuracy in \teff\ and \logg.  We then rejected the stars whose ages are determined from stellar models with temperatures, metallicites or surface gravities deviating from input data by more than 300K, 0.3 dex, and 0.3 dex, respectively (i.e., \texttt{AgeInOut $\neq$ "WARN\_diff\_inout"}, see \citealt{queiroz2023}). We note that the main conclusion of this work does not change if we replace [$\alpha$/M] with [Mg/Fe] for the target selection. 

Subsequently, we pruned the YAR and OAR samples by preserving stars for which the square root of their vertical actions exceeded 2.5 kpc km s$^{-1}$ (see Fig.~\ref{fig:sampleselection}b). Here, the vertical actions were adopted from \href{https://www.sdss4.org/dr17/data_access/value-added-catalogs/?vac_id=the-astronn-catalog-of-abundances,-distances,-and-ages-for-apogee-dr17-stars}{the \texttt{astroNN} catalog} and they were calculated with a method detailed in \citet{mackereth2018}. They serve as a kinematic age proxy, and higher vertical actions correspond to older ages. This kinematic filtering was meant to minimize any contaminants from the young, thin disk. We anticipated a minimal fraction of YAR contaminants, if exist, in the YAP population due to the significantly higher number of YAP stars (see Fig.~\ref{fig:sampleselection}a). Again, we note that the main conclusion of this work remains unchanged if we do not apply this additional cut in kinematics. After necessitating the availability of \vsini\ from APOGEE DR17, the entire selection procedure finally yields 335 YAR stars, 2930 OAR stars, and 43,141 YAP stars. 

\begin{figure}
\includegraphics[width=\columnwidth]{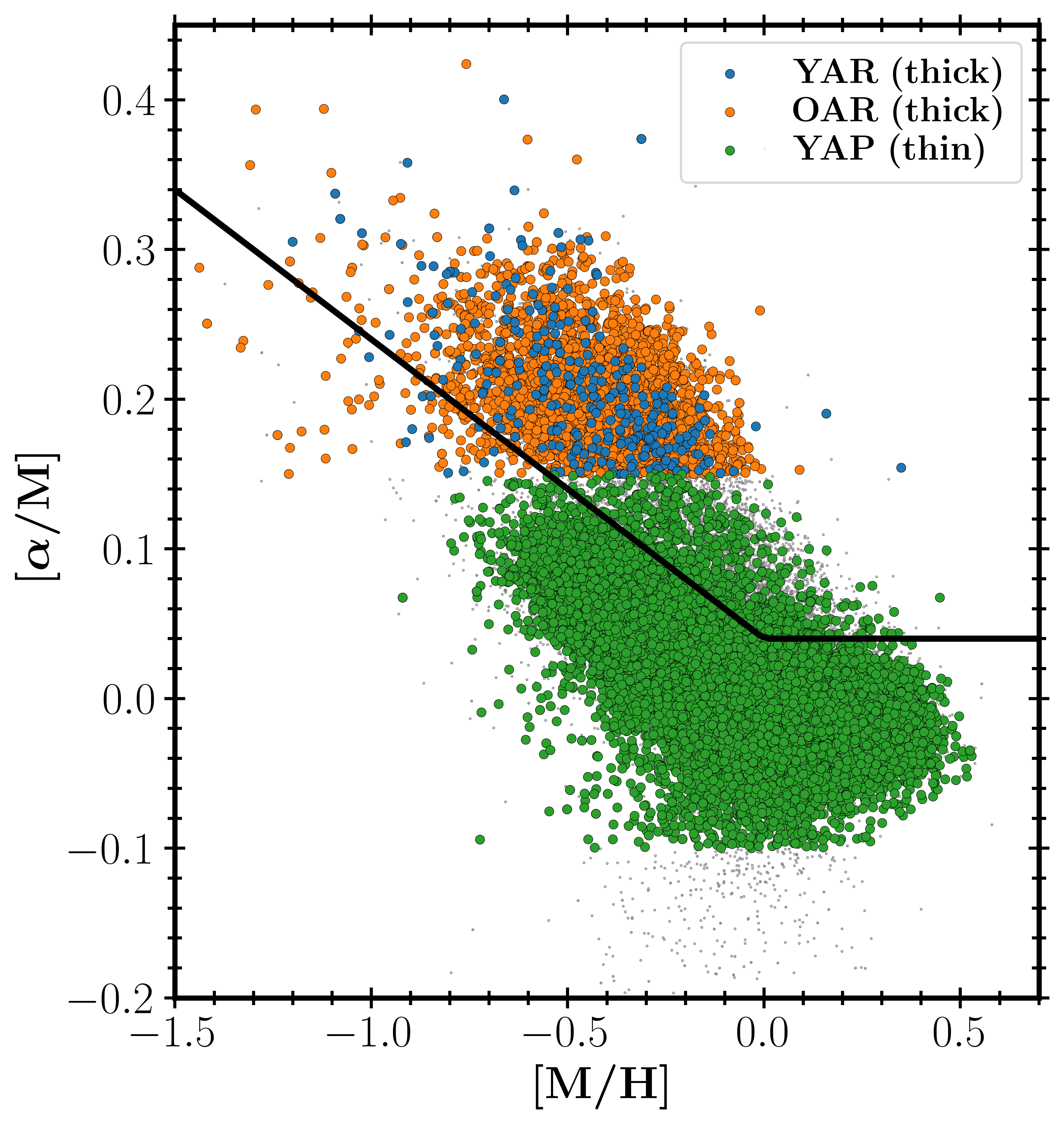}
\caption{Tinsley--Wallerstein diagram (aka [$\rm \alpha$/M] versus [M/H]) of  YAR (blue), OAR (orange), and YAP (green) stars. The definition of each sample may be found from Fig.~\ref{fig:sampleselection}. The black lines represent a commonly used method for defining YAR stars (above the lines, \citealt{miglio2021}), indicating our criterion is conservative. The small grey dots indicate the rest stars unclassified. The [M/H] and [$\alpha$/M] values are adopted from APOGEE DR17.
}\label{fig:twdiagram}
\end{figure}

\begin{figure}
\includegraphics[width=\columnwidth]{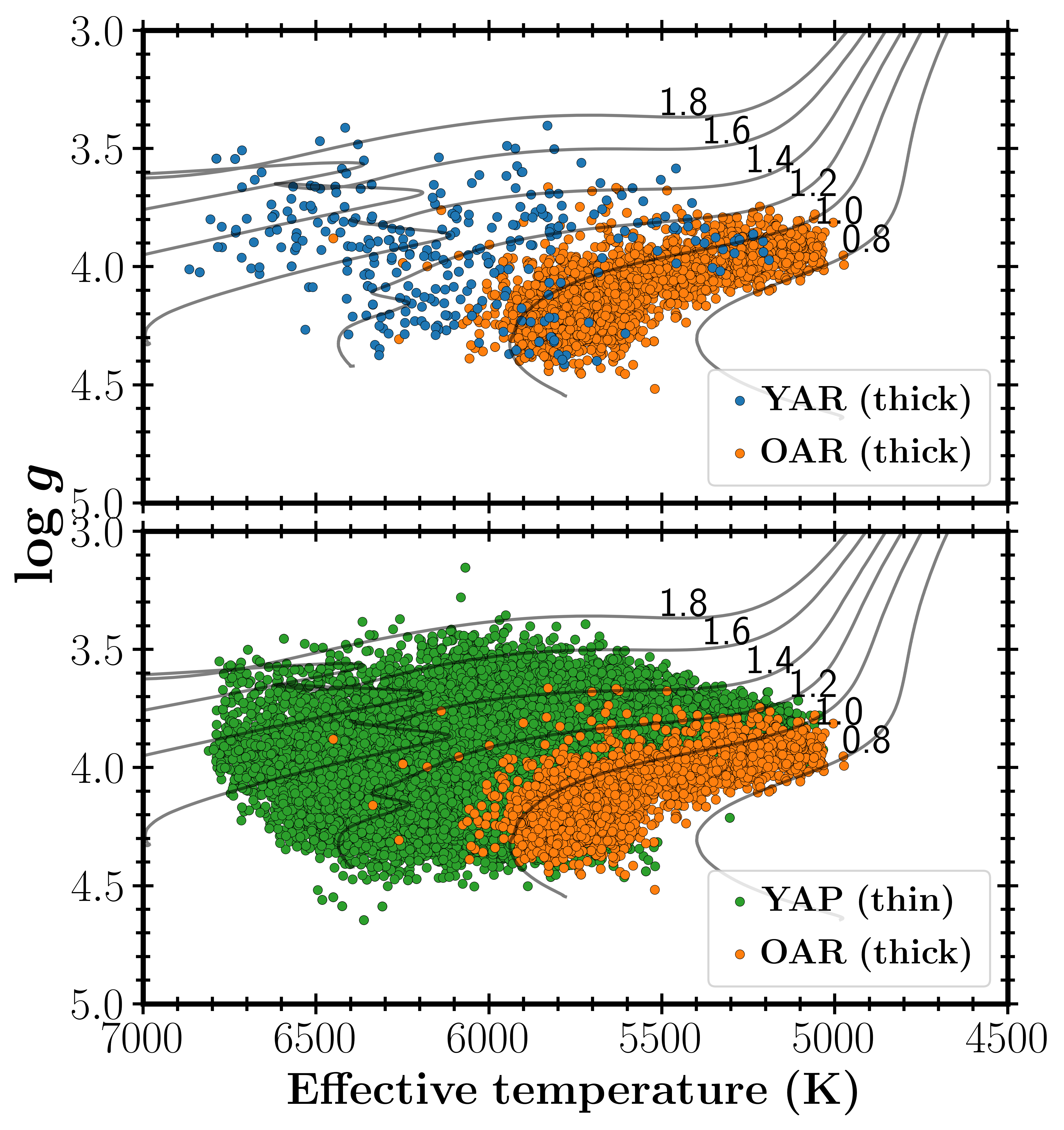}
\caption{Kiel diagram (\logg\ vs [\teff]) of YAR (blue), OAR (orange), YAP (green) stars. The PARSEC evolutionary tracks of solar metallicity are overplotted, with masses ranging from 0.8 to 1.8 \msun\ in steps of 0.2 \msun\ from right to left (see labels near tracks).  The stars shown here are the same as in Fig.~\ref{fig:twdiagram}. The \teff\ and \logg\ values are adopted from APOGEE DR17.
}\label{fig:kiel}
\end{figure}

It is worth noting that our selection scheme of YAR stars is conservative in terms of their chemistry and kinematics. Regarding chemistry, according to the selection criterion of \citet{miglio2021}, young $\alpha$--rich can have \alpham\ ratios down to 0.04 at solar metallicity (see the Tinsley--Wallerstein diagram in Fig.~\ref{fig:twdiagram}). As to kinematics, the thick disk, to which YAR stars belong, can have $\rm \sqrt{J_{z}}$ down to a few kpc km s$^{-1}$\citep{ciuca2023}. 

\begin{figure*}
\includegraphics[width=\textwidth]{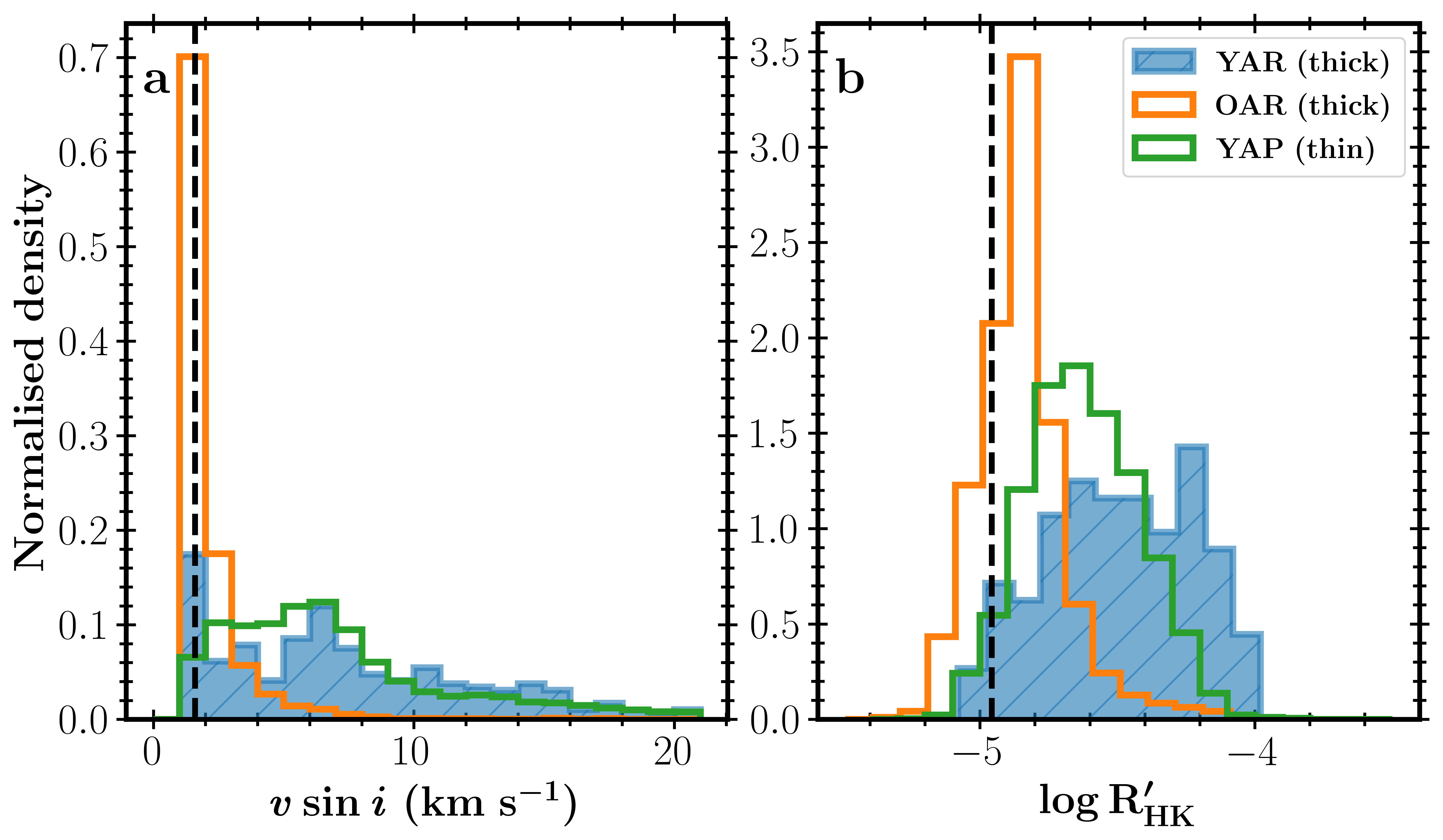}
\caption{Histograms of \vsini\ adopted from APOGEE DR17 and the activity proxy \logrhk\ measured from Ca II H \& K lines of LAMOST spectra for YAR (blue), OAR (orange), YAP (green) stars. The vertical dashed lines denote solar \vsini\ (1.6 km~s$^{-1}$, \citealt{pavlenko2012}) and \logrhk\ (-4.96). The target selection scheme of the three populations can be found from Fig.~\ref{fig:sampleselection}.}\label{fig:vsiniactivity}
\end{figure*}

Fig.~\ref{fig:kiel} shows the three samples in the Kiel diagram. We can see the YAR and YAP stars have similar distributions and are globally more massive and hotter than the OAR stars. None of these selected stars have begun ascending the red giant branch, where the first dredge--up takes place. Thus, the chemistries of these stars are not directly altered by the chemical mixing associated with this process.

\section{Activity proxy measurements: $\log R^{\prime}_{\mathrm{HK}}$} \label{sec:activityindex}
We employed the method in Yu et al. (2024, in prep.) to calculate a chromospheric activity index, \logrhk, which is based Ca II H \& K lines (3968 \AA\ and 3934 \AA, respectively) of \href{http://www.lamost.org/dr9/v2.0/}{LAMOST DR9v2.0 } low--resolution spectra \citep{cui2012}. To locate Ca II H \& K lines, we determined radial velocities from \halpha\ ($\lambda_0=6563 $ \AA\ in air) and Na D lines ($\lambda_0=5893 $ \AA\ in air) in the optical that are usually of higher single--to--noise ratios (SNRs, see a similar method in \citealt{gehan2022}). Specifically, we fitted a composite model comprised of a line and a Gaussian to \halpha\ and Na D lines, and the central Gaussian wavelength $\lambda$ for each line was used to calculate radial velocity. The final radial velocity estimate was obtained from the strongest line with an absorption depth of at least 6 per cent.

\begin{figure*}
\includegraphics[width=\textwidth]{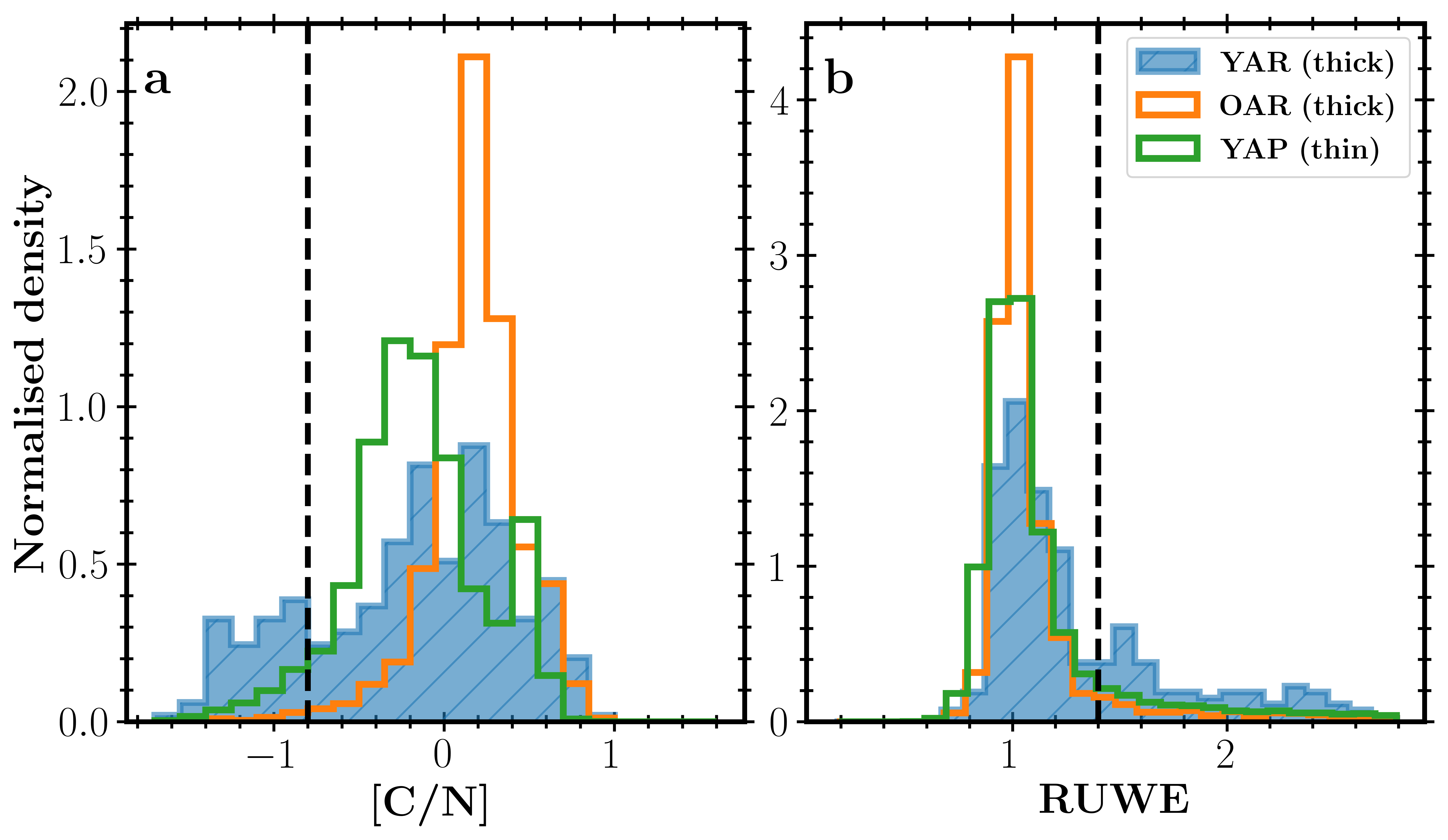}
\caption{Histograms of [C/N] ratios adopted from APOGEE DR17 and Gaia DR3 RUWE data for YAR (blue), OAR (orange), YAP (green) stars. In the YAR population, approximately 21 per cent of stars exhibit [C/N] $<-0.8$ (left dashed vertical line), and 45 per cent have RUWE $>$ 1.4 (right dashed vertical line). The target selection scheme of the three populations can be found from Fig.~\ref{fig:sampleselection}.}\label{fig:cnruwe} 
\end{figure*}

After switching spectra to rest wavelengths using radial velocities, integrated emission line fluxes in H and K bandpasses were calculated using 4.36 \AA\ FWHM triangular windows. Compared with the 1.36 \AA\ window used by \citet{gomesdasilva2021} for analysing HARPS spectra with a resolution power of 115,000, our window size was selected to be wider, in order to ensure that all the emission in Ca II H \& lines are captured in LAMOST spectra with a lower average resolution of 1800. Continuum fluxes for R and V bandpasses (4001 \AA\ and 3901 \AA) were determined using 20 \AA\ rectangular windows. Following \cite{karoff2016}, the $S$ index, which is later used for calculating \logrhk, was computed as below:

\begin{equation}
S = \alpha \times 8 \times \frac{4.36~\text{\r{A}}}{20~\text{\r{A}}} \times \frac{H+K}{R+V},
\end{equation}
where the $S$--index calibration factor $\alpha=1.8$. The uncertainty of the $S$ index was estimated using the method proposed by \citet{karoff2016}, where $\log{\sigma(S)}=-\log(\overline{S/N})-0.5$, with $\overline{S/N}$ being the average SNR (over 3841--4061 \AA) obtained from LAMOST spectrum files. When multiple epoch spectra are available, the mean $S$ index value and uncertainty were used.

To standardize $S$ index measurements, we calibrated them by applying our method to a sample of 1747 \textit{Kepler} stars with known $S$ index values anchored to the Mount Wilson scale by \citet{karoff2016}. This calibration ensured consistency despite wider triangular H and K bandpasses ($4.36$ \AA) and lower spectrum resolutions. 

We then calculated \logrhk\ from the $S$ index. Specifically, the \primerhk\ metric represents the ratio of chromospheric fluxes in Ca II H \& K lines to the bolometric flux. Thus, it involves subtracting the photospheric flux (\rhkphot) from the chromospheric surface flux (\rhk). Following \citet{noyes1984}, we computed $R_{\textrm{HK}}$ as:
\begin{equation}
R_{\textrm{HK}} = 1.34 \times\ 10^{-4}~C_{\textrm{cf}}~S.
\end{equation}
Here, $C_{\textrm{cf}}$ is a color--dependent conversion factor that converts the $S$ index to the surface flux, and hence enables us to investigate the magnetic activity for stars with different spectral types. The calculation of $\log C_{\textrm{cf}}$ for dwarf stars with $0.3<(B-V)<1.6$ adheres to the scheme of \citet{rutten1984}:
\begin{equation}
\log C_{\textrm{cf}} = 0.25(B-V)^3 - 1.33(B-V)^2 + 0.43(B-V) + 0.24
\end{equation}
The determination of \rhkphot\ follows the approach of \citet{hartmann1984} and \citet{noyes1984}:
\begin{equation}
\log \rhkphot = -4.898 + 1.918 (B-V)^2-2.893(B-V)^3.
\end{equation}

We calculated $(B-V)$ from spectroscopic $T_{\rm eff}$ provided by LAMOST DR9v2.0. Adopting values from \citet{huang2015} for 134 dwarf stars with 3100~\textrm{K}$<$\teff$<$9700~\textrm{K}, we derived the quadratic polynomial:
\begin{equation}
(B-V) = 3.964 * (T_{\text{eff}}/10^4)^2 - 7.709 * (T_{\text{eff}}/10^4) + 3.738.
\end{equation}
Finally, we obtained $\primerhk=\rhk-\rhkphot$ and its logarithmic scale, \logrhk. As an example, if taking $(B-V)=0.66$ and $S=0.171$ for the Sun \citep{noyes1984}, we obtain \logrhk=$-4.96$.
The application of this methodology led to the determination of \logrhk\ measurements for 112 YAR stars, 944 OAR stars, and 11,742 YAP stars in common between the three pre-selected polulations and LAMOST DR9v2.0.

\section{Tracing Binary footprint in young--$\alpha$ rich stars}
Fig.~\ref{fig:vsiniactivity}a illustrates the \vsini\ distributions of the YAR, YAP, OAR stars. Despite being located in the old, thick disk, the YAR stars exhibit faster rotation compared with the OAR stars and are similar to the YAP stars in the young, thin disk. The median \vsini\ values are 7.1, 6.6, and 1.5 km~s$^{-1}$ for the YAR, YAP, and OAR stars, respectively. The fast rotation nature in the YAR stars suggests they should be spun up by gaining angular momentum due to mass transfer or stellar mergers, since they would otherwise chemically be expected to rotate similarly to the OAR stars. The peak consisting of $\sim7$ per cent YAR stars at \vsini\ $<2$ km~s$^{-1}$, primarily characterized by weak activity, is likely subject to mis-classification. 

In addition to \vsini, Fig.~\ref{fig:vsiniactivity}b illustrates the  chromospheric activity, \logrhk, of the three samples. One can see that the YAR stars are globally the most active group. Specifically, the median \logrhk\ values are -4.49, -4.64, and -4.86 for the YAR, YAP and OAR stars, respectively. To understand this, we recall that rapid rotation is the generator of strong magnetic fields in the interior of the star, which emerges on the surface and manifests as magnetic activity in the photosphere, chromosphere, and corona \citep[see][for a review]{brun2017}. This activity saturates at highest rotational rates (e.g., \citealt{wright2011}). As such, Fig.~\ref{fig:vsiniactivity}b suggests that the YAR stars spin rapidly, at rotation rates higher than the OAR stars. This is consistent with the result demonstrated in Fig.~\ref{fig:vsiniactivity}a in the sense that the YAR stars are fast rotators. We note that the YAR stars consistently represent the rapid-rotating and most active population when restricting the three samples to the same \teff\ bin between 5600 and 5800 K. Given \vsini\ and \logrhk\ are two different metrics and measured with different data sets (APOGEE spectra versus LAMOST spectra), we argue that Fig.~\ref{fig:vsiniactivity} serves as \highlight{new} evidence in support of the hypothesis that binary evolution \highlight{is} the origin of YAR stars. 

\highlight{We examined the relationship between age and activity for the YAP and YAR samples, and observed that YAR stars appear to exhibit slightly higher levels of activity compared to YAP stars at fixed ages. However, this observation requires further confirmation with a larger sample size. Prior studies have suggested a potential, albeit weak, positive correlation between activity and metallicity \citep[see][and references therein]{see2023}. Nevertheless, this correlation cannot account for the observed difference in activity between YAR and YAP stars illustrated in Fig.~\ref{fig:vsiniactivity}b, where YAR stars with lower metallicities demonstrate collectively stronger activity.} 

One might suspect that if the $\alpha$ abundances of the YAR stars were inaccurate and the selected YAR and YAP stars belonged to the same population (see Fig.~\ref{fig:sampleselection}a), the \vsini\ and \logrhk\ distributions would appear similar. However, this scenario is unlikely, because Fig.~\ref{fig:sampleselection}b independently indicates that the \highlight{YAR} and OAR populations are similar in terms of their high vertical actions, representative of thick--disk stars.

We then examined [C/N] ratios to trace the yield of the binary evolution in  MSTO--SGB stars, where the ratios may be altered due to the accreting material originating from the companion in the system \citep{izzard2018}. Fig.~\ref{fig:cnruwe}a shows that the [C/N] ratios of both YAR and OAR stars peak at $\simeq$0.19, reflecting the global chemistry property of the thick disk. However, it is noteworthy that some YAR stars possess exceedingly low [C/N] ratios, manifesting a second peak at [C/N]$<-0.8$. Among these stars, we identified that 49.2 per cent have RUWE values $>$ 1.4. Since the chemical mixing associated with the first dredge--up does not occur in these MSTO--SGB stars, the presence of these reduced [C/N] ratios suggest they could stem from mass transfer from red--giant companions that have evolved to fainter stars, such as white dwarfs (e.g., \citealt{hekker2019, zhang2021a, bufanda2023, grisoni2023}).

\begin{figure*}
\includegraphics[width=0.9\textwidth]{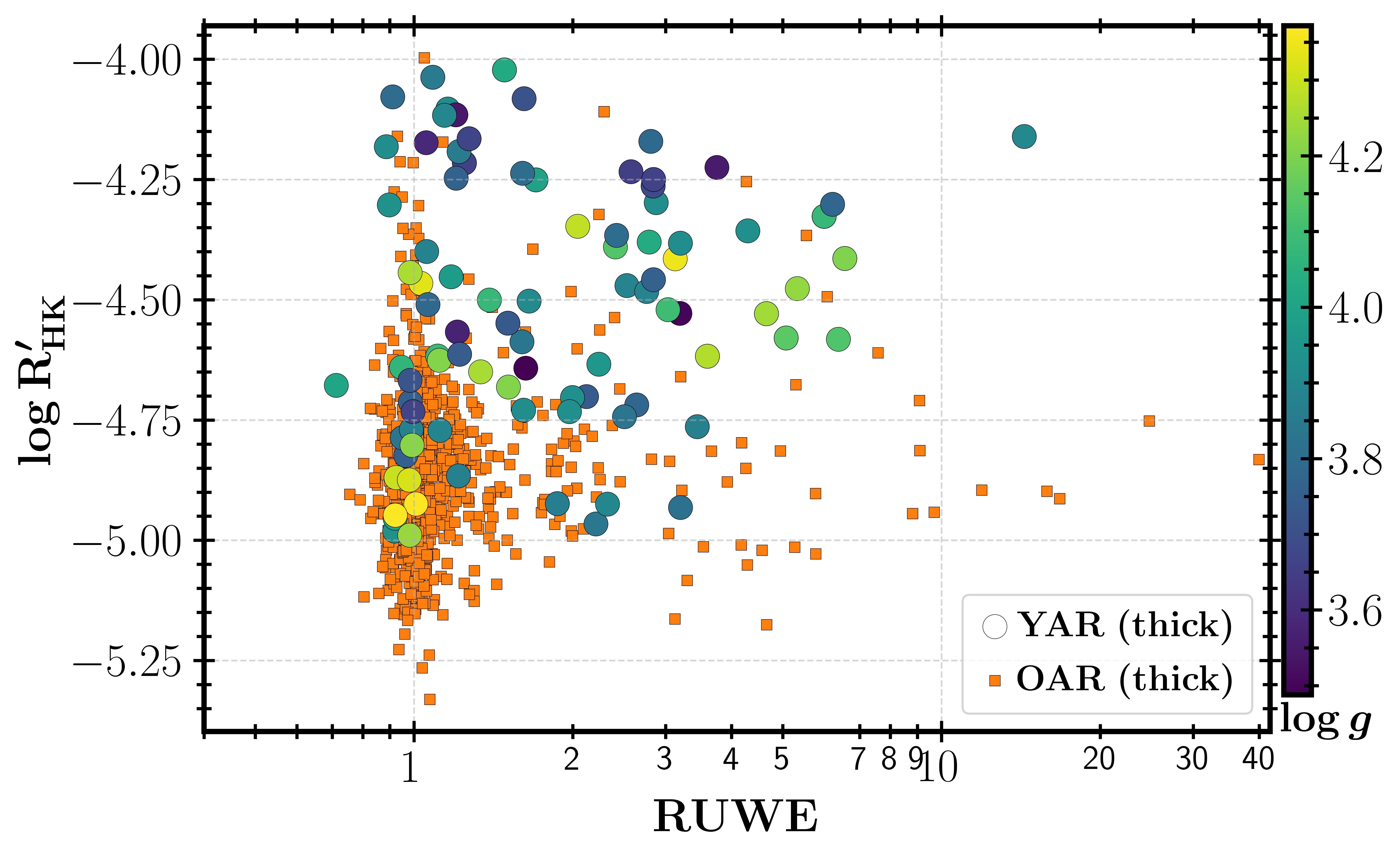}
\caption{Relationship between Gaia RUWE and the magnetic activity index of \logrhk. Yellow squares indicate the OAR population, while circles represent the YAR stars colour--coded by \logg\ adopted from APOGEE DR17.}\label{fig:ruwelogrhk}
\end{figure*}

We also investigated Gaia DR3 RUWE data (Fig.~\ref{fig:cnruwe}b). On one hand, our observations reveal that the RUWE distributions among the three samples all peak at $\sim$1.02. This similarity is probably expected due to the short, dynamical timescales associated with merger processes. On the other hand, we found that 45 per cent of the YAR stars have RUWE $>$ 1.4, as compared with 11 per cent of the OAR and 21 per cent of the YAP stars. This result supports the binary interaction hypothesis, potentially mass transfer due to detectable binarity with RUWE. It is noteworthy that only 7 per cent (2/28) YAR red giants in \citet{jofre2023} exhibit RUWE $>$ 1.4. This reduced proportion may suggest increased difficulty in identifying binaries with red giants. A larger sample of YAR red giants will be valuable to further investigate this possibility.

Interestingly, upon inspecting the RUWE measurements of the most active YAR stars (top 25 percentile), we found that their median RUWE is 1.2, even lower than the median RUWE (2.8) of the YAR stars with \logrhk\ between the 50th and 75th percentiles (see Fig~\ref{fig:ruwelogrhk}). In contrast, the median \logrhk\ of the YAP stars with high RUWE values exceeding 1.4 is -4.4, similar to the typical activity level of the entire YAR population. The absence of a clear correlation between \logrhk\ and RUWE suggests that they potentially trace binary evolution in different dimensions, perhaps mergers and mass transfer, respectively. In addition, 
Fig~\ref{fig:ruwelogrhk} shows that more evolved YAR stars (with lower \logg) seem to be globally more active (stars at the top left corner), which is different from single stars that spin down due to magnetic braking \citep[e.g., see][]{gomesdasilva2021}. This observation suggests a potential influence of binary interactions on the magnetic activity of YAR stars.  

\highlight{We note that incorporating Gaia parallaxes with high RUWE values for isochrone fitting might introduce biases in derived stellar parameters. To address this concern, we investigated the relationship between RUWE and $\delta$\teff, where $\delta$\teff\ represents the difference between input and output \teff\ in the \texttt{StarHorse} catalog. Our analysis revealed that RUWE shows no correlation. Similarly, we observed a comparable relationship between RUWE and $\delta$\logg, where $\delta$\logg\ denotes the difference between input and output \logg. These results indicate that Gaia parallaxes with high RUWE do not significantly bias the output \teff\ or \logg\ values, which were used for the identification of the \mbox{MSTO-SGB} stars.}

\section{conclusion}
We investigated the youth origin of YAR stars in the evolutionary phase of the MSTO--SGB, \highlight{whose} ages can be precisely measured from isochrone fitting. We selected a sample of YAR stars and two reference samples of OAR and YAP stars using stellar ages and [$\rm \alpha$/M] ratios (see Section~\ref{sec:sample} and Fig.\ref{fig:sampleselection}a). We adopted stellar age measurements from the \texttt{StarHorse} catalog that are based on atmospheric parameters provided by APOGEE DR17 \citep{queiroz2023}. We then pruned the YAR star sample by including the targets with the square root of vertical actions exceeding 2.5 kpc km s$^{-1}$ (Fig.\ref{fig:sampleselection}b). This was to ensure the YAR star sample representative of the Galactic thick disk.

Our primary motivation was to test the binary interaction hypothesis that YAR stars spin rapidly and are magnetically active due to mass transfer or stellar mergers. We analyzed \vsini\ estimates adopted from APOGEE DR17 and the activity indicator of \logrhk\ determined with Ca II H \& K lines of LAMOST DR9v2.0 spectra. Our findings show that YAR stars are indeed faster rotators and more magnetically active than OAR stars in the thick disk (Fig.~\ref{fig:vsiniactivity}). Additionally, we observed that a subset of YAR stars have exceptionally low $[\text{C}/\text{N}]$ ratios and exceedingly high Gaia RUWE values (Fig.~\ref{fig:cnruwe}). This can point to mass transfer from red giant companions rather than mergers, otherwise higher luminosities due to mergers would be incompatible with the MSTO--SGB phase. We also found the absence of a clear correlation between magnetic activity and Gaia RUWE (Fig.~\ref{fig:ruwelogrhk}). This implies that magnetic activity and Gaia RUWE can serve as valuable probes for binary evolution along distinct dimensions, perhaps preferentially for mergers and mass transfer, respectively.

The interpretation of the origins of several types of chemically peculiar stars has been suggested to binary evolution. In addition to YAR stars, other examples encompass \mbox{lithium--rich} red giants \citep[e.g.][]{casey2019}, carbon--enhanced metal--poor stars \citep[e.g.][]{starkenburg2014}, and blue stragglers \citep[e.g.][]{geller2011}. Independent of the metrics, such as RV variability, chemical abundances, and Gaia RUWE, the examination of stellar rotation and magnetic activity offer a unique perspective for testing the binary evolution hypothesis. Further investigations in this regard would be valuable and could contribute to a deeper understanding of the origins of these peculiar stars.

\section*{Acknowledgements}
Parts of this research were conducted by the Australian Research Council Centre of Excellence for All Sky Astrophysics in 3 Dimensions (ASTRO 3D) through project number CE170100013. JY expresses gratitude for the Research Fellowship provided by the School of Computing at the Australian National University. J.Y. acknowledges the Joint Research Fund in Astronomy (U2031203) under a cooperative agreement between the National Natural Science Foundation of China (NSFC) and the Chinese Academy of Sciences (CAS). IC is grateful for the Joint Jubilee Fellowship at the Australian National University. YST acknowledges financial support from the Australian Research Council through DECRA Fellowship DE220101520. SJM was supported by the Australian Research Council (ARC) through Future Fellowship FT210100485.

\section*{Data Availability}
We utilised publicly available data for this study. To facilitate the reproduction of the main results presented in this work, the data supporting this article will be provided upon reasonable request to the corresponding author.

\bibliographystyle{mnras}
\bibliography{references.bib} 


\bsp 
\label{lastpage}
\end{document}